\documentclass[10pt,reqno]{amsart}
  \usepackage{geometry}
\usepackage{MnSymbol} 

\newcommand{\be}{\begin{equation}}
\newcommand{\ee}{\end{equation}}
\newcommand{\ba}{\begin{eqnarray}}
\newcommand{\ea}{\end{eqnarray}}
\newcommand{\baa}{\begin{eqnarray*}}
\newcommand{\eaa}{\end{eqnarray*}}
\newcommand{\bb}{\begin{thebibliography}}
\newcommand{\eb}{\end{thebibliography}}

\newcommand{\bi}[1]{\bibitem{#1}}
\newcommand{\lab}[1]{\label{#1}}

\newcounter{my}
\newcommand{\he}%
   {\stepcounter{equation}\setcounter{my}%
   {\value{equation}}\setcounter{equation}0%
   }%
\newcommand{\she}%
   {\setcounter{equation}{\value{my}}%
    }%

\newtheorem{theorem}{Theorem}[section]

\theoremstyle{definition}

\newtheorem{remark}[theorem]{Remark}

\numberwithin{equation}{section}

\begin{document}

\title[Classical Heun observables and elliptic solvability]
{Classical Heun observables and elliptic solvability}

\author{Luc Vinet}
\author{Alexei Zhedanov}

\address{Centre de recherches math\'ematiques,
Universit\'e de Montr\'eal, P.O.\ Box 6128, Centre-ville Station,
Montr\'eal (Qu\'ebec), H3C~3J7}

\address{Euler International Mathematical Institute, Saint Petersburg, Russia}

\begin{abstract}
We introduce a classical analog of the algebraic Heun operator associated
with a classical Leonard pair.  Given two observables $X$ and $Y$ satisfying
the classical counterpart of the Askey--Wilson relations, we define a
\emph{classical Heun observable} $W$ as the most general bilinear combination
of $X$, $Y$, and their Poisson bracket.  We prove that, when $W$ is taken as
Hamiltonian, the dynamics of X and Y is governed by quartic differential equations and, generically, by elliptic functions of second order. This result provides a universal algebraic mechanism
transforming the elementary dynamics associated with classical Leonard pairs
into elliptic dynamics, and yields an algebraic explanation of a classical
observation of Manning on the connection between the Heun equation and
elliptic solvability.  The construction is illustrated on three examples: an
extension of the P\"oschl--Teller system, the Zhukovsky--Volterra gyrostat,
and a relativistic $A_1$ model related to the classical Askey--Wilson algebra.
\end{abstract}

\keywords{}

\maketitle

\bigskip

\section{Introduction}
\setcounter{equation}{0}

One of the remarkable early observations on exactly solvable systems is due
to Manning~\cite{Manning}.  In a 1935 paper that has not received the
attention it deserves, he pointed out a striking parallel between classical
and quantum notions of solvability.  His first observation is that the
quantum systems whose Schr\"odinger equation reduces to the hypergeometric
equation are precisely those whose classical Hamilton--Jacobi equation can be
solved in terms of elementary functions.  His second observation is that,
whenever the classical dynamics is described by elliptic functions, the
corresponding Schr\"odinger equation belongs to the Heun class.

The first part of this correspondence was given a satisfactory algebraic
interpretation in~\cite{GLZ_Annals}: both the classical and quantum systems
are governed by the same hidden quadratic Jacobi algebra, which explains why
elementary functions appear naturally in both settings.  The second part ---
the link between Heun equations and elliptic classical dynamics --- has
remained much less transparent. The purpose of the present paper is to
supply this missing algebraic explanation.
Within the framework of classical Leonard pairs and
their associated Heun observables, we also establish
a reverse mechanism: the Heun deformation of a system
with elementary dynamics naturally generates elliptic
classical dynamics.

Our starting point is the notion of \emph{algebraic Heun operator} introduced
in~\cite{GVZ_Heun, GVZ_band}.  Let $X$ and $Y$ be a bispectral pair of
operators.  The algebraic Heun operator is the most general bilinear
combination of $X$ and $Y$:
\begin{equation}
W=
\tau_1(XY+YX)
+\tau_2[X,Y]
+\tau_3X
+\tau_4Y
+\tau_0\mathcal I.
\label{alH}
\end{equation}
This construction originated from the observation that, when $X$ is the
multiplication operator and $Y$ is the hypergeometric operator, the
operator~$W$ coincides with the generic Heun operator~\cite{GVZ_Heun}.  It
was subsequently shown in~\cite{GVZ_band} that algebraic Heun operators also
provide a natural framework for the band--time limiting problem. The Heun operator has also been studied as a Hamiltonian in its own right; see for instance \cite{Tur_Heun, Tur_quasi}.

The central idea of the present paper is to construct a classical counterpart
of~\eqref{alH}.  Starting from a \emph{classical Leonard pair} $(X,Y)$, we
introduce the \emph{classical Heun observable}~$W$ by replacing the
commutator in~\eqref{alH} with the Poisson bracket.  Our main result states
that, on every energy surface $W = \mathrm{const}$, the variables $X$ and $Y$
satisfy
\[
\dot X^2=\mathcal P_4(X),
\qquad
\dot Y^2=\widetilde{\mathcal P}_4(Y),
\]
where $\mathcal P_4$ and $\widetilde{\mathcal P}_4$ are quartic polynomials.
In the generic non-degenerate case, both variables therefore evolve as
elliptic functions of second order.

The passage from a Leonard observable to its Heun deformation may thus be
viewed as a universal algebraic mechanism transforming elementary dynamics
into elliptic dynamics.  Our construction applies uniformly to all classical
Leonard pairs, is independent of their specific realization, and provides a broad generalization of Manning's observation
and reveals an algebraic mechanism that operates in the
reverse direction: starting from a classical Leonard pair,
the associated Heun deformation produces elliptic dynamics.  In addition, our framework encompasses not only
Schr\"odinger-type operators but the more general operators of Askey--Wilson
type and their classical analogs.

From a conceptual viewpoint, the present work establishes
a hierarchy

\[
\text{classical Leonard pair}
\longrightarrow
\text{elementary dynamics},
\]

\[
\text{classical Heun observable}
\longrightarrow
\text{elliptic dynamics},
\]
which may be viewed as the classical counterpart of the
transition from hypergeometric to Heun structures in
quantum mechanics.

The paper is organized as follows.  Section~2 introduces the classical Heun
observable associated with a classical Leonard pair and derives the universal
quartic equations governing the evolution of $X$ and $Y$.  Section~3
illustrates the construction on three physical examples.  The first provides
a direct connection with Manning's original discussion via a Heun extension of
the P\"oschl--Teller system.  The second identifies the Zhukovsky--Volterra
gyrostat as a Heun deformation of a classical Leonard pair on the
Lie--Poisson algebra $\mathfrak{su}(2)$.  The third involves a relativistic
$A_1$ model associated with the classical Askey--Wilson algebra.  Concluding
remarks are gathered in Section~4.

\section{Classical analog of the algebraic Heun operator}
\setcounter{equation}{0}

Under the standard classical--quantum correspondence, an operator $X$
becomes an ordinary observable, the anticommutator $XY+YX$ becomes $2XY$,
and the commutator becomes the Poisson bracket $[X,Y] \to \{X,Y\}$.  In the
one-dimensional case the observables $X(q,p)$ and $Y(q,p)$ depend on
canonical coordinates $q$ and $p$, and the Poisson bracket is the Jacobian
\be
\{X,Y\} = \frac{D(X,Y)}{D(q,p)} \equiv X_q Y_p - X_p Y_q, \lab{PB_def}
\ee
where $F_s = \partial_s F$.  Following this prescription, we define the
\emph{classical Heun observable}
\be
W = \tau_1 XY + \tau_2 Z + \tau_3 X + \tau_4 Y + \tau_0, \lab{cl_W}
\ee
where
\be
Z = \{X,Y\} \lab{Z_XY}
\ee
is the Poisson bracket of $X$ and $Y$, and where the $\tau_i$ are real
constants independent of time and of the canonical variables.  We shall also
refer to $W$ as the \emph{classical Heun pencil}, to emphasize that it is a
linear combination with arbitrary coefficients $\tau_i$.
The construction becomes particularly interesting when
the observables $X$ and $Y$ form a classical Leonard pair.
Algebraically, a classical Leonard pair is a pair of
observables satisfying the relations \cite{ZK}

\be
\{X,Z\} = \tfrac{1}{2} \Phi_Y(X,Y),
\qquad
\{Z,Y\} = \tfrac{1}{2} \Phi_X(X,Y), \lab{cl_alg}
\ee
where $\Phi(X,Y)$ is a polynomial of degree at most two in each variable,
\be
\Phi(X,Y) = \sum_{i,j=0}^2 \alpha_{ij} X^i Y^j, \lab{Phi_def}
\ee
and $\Phi_X, \Phi_Y$ denote its partial derivatives.  We also introduce the
auxiliary polynomials
\be
U_i(x) = \alpha_{2i} x^2 + \alpha_{1i} x + \alpha_{0i},
\quad
V_i(x) = \alpha_{i2} x^2 + \alpha_{i1} x + \alpha_{i0},
\quad i = 0,1,2, \lab{UV_pol}
\ee
so that $\Phi$ admits the two factored forms
\ba
\Phi(X,Y) &= U_2(X) Y^2 + U_1(X) Y + U_0(X), \nonumber \\
\Phi(X,Y) &= V_2(Y) X^2 + V_1(Y) X + V_0(Y). \lab{Phi_UV}
\ea

\begin{remark}
Relations~\eqref{cl_alg} define the classical counterpart of the
Askey--Wilson algebra.  A pair $(X,Y)$ satisfying these relations is called a
\emph{classical Leonard pair}~\cite{ZK}, providing the natural phase-space
analog of Leonard pairs and bispectral systems in the theory of orthogonal
polynomials.
\end{remark}

A key structural feature of every CLP is that the observable
\be
Q(q,p) = Z^2 - \Phi(X,Y) \lab{Q_cas}
\ee
Poisson-commutes with both $X$ and $Y$~\cite{ZK},
\be
\{Q,X\} = \{Q,Y\} = 0. \lab{Q_0}
\ee
Hence $Q$ is a constant of motion that can be absorbed into the free term
$\alpha_{00}$ of~$\Phi$, yielding the fundamental identity
\be
Z^2 = \Phi(X,Y). \lab{Z^2_Phi}
\ee

\subsection*{Elementary dynamics of a CLP}

To set the stage for the main result, we recall how the choice $H = X$ (or
$H = Y$) leads to elementary dynamics~\cite{GLZ_Annals, ZK}.  With $H = X$,
Hamilton's equation gives
\be
\dot Y = \{Y, X\} = -Z, \lab{dot_Y_X=H}
\ee
and combining with~\eqref{Z^2_Phi} yields
\be
\dot Y^2 = \mu_2(Y), \lab{dot_Y_X=H_2}
\ee
where $\mu_2$ is a polynomial of degree at most two in $Y$ with coefficients
depending only on $X$.  The general solution is therefore an elementary
function of time:
\be
Y(t) = \xi_1 e^{\omega t} + \xi_2 e^{-\omega t} + \xi_0, \lab{Y_exp}
\ee
or, in the degenerate case, a polynomial $Y(t) = \xi_2 t^2 + \xi_1 t +
\xi_0$.  The companion variable $Z(t) = -\dot Y(t)$ is elementary as well.
By the symmetry of the CLP in $X$ and~$Y$, the same conclusion holds with
$H = Y$, giving
\be
\dot X^2 = \nu_2(X) \lab{dot_X_Y=H_2}
\ee
for a second-degree polynomial $\nu_2$.  This \emph{mutual integrability}
property is the hallmark of a classical Leonard pair.

\subsection*{Elliptic dynamics of the Heun observable}

We now turn to the Hamiltonian system defined by the Heun observable $W$.
Hamilton's equations of motion are
\be
\dot X = \{X, W\}, \quad \dot Y = \{Y, W\}. \lab{dot_XY}
\ee
Using the CLP relations~\eqref{cl_alg}, these Poisson brackets evaluate to
\be
\{X,W\} = \tau_1 XZ + \tfrac{\tau_2}{2}\, \Phi_Y(X,Y) + \tau_4 Z,
\lab{PB_XW}
\ee
\be
\{Y,W\} = -\tau_1 YZ - \tfrac{\tau_2}{2}\, \Phi_X(X,Y) - \tau_3 Z.
\lab{PB_YW}
\ee

The key step is to eliminate $Y$ and $Z$ from the three
equations~\eqref{cl_W}, \eqref{Z^2_Phi}, and~\eqref{PB_XW}.  This yields a
relation involving only $X$, $W$, and $\{X,W\}$:
\be
\{X,W\}^2 = \pi_2(X)\,W^2 + \pi_3(X)\,W + \pi_4(X), \lab{PBWX_pol}
\ee
where the polynomials $\pi_i$ have degree at most $i$ in $X$.  Their
explicit forms are
\be
\pi_2(x) = U_2(x),
\quad
\pi_3(x) = (\tau_1 x+\tau_4)\,U_1(x) - 2(\tau_3 x+\tau_0)\,U_2(x),
\lab{pi_23}
\ee
\be
\pi_4(x) = U_0(x)(\tau_1 x+\tau_4)^2
           - U_1(x)(\tau_1 x+\tau_4)(\tau_3 x+\tau_0)
           + \tfrac{\tau_2^2}{4}\bigl[U_1^2(x)-4\,U_0(x)\,U_2(x)\bigr].
\lab{pi_4}
\ee
An entirely analogous computation, replacing $U_i \to V_i$ and
$\tau_3 \leftrightarrow \tau_4$, gives
\be
\{Y,W\}^2 = \tilde\pi_2(Y)\,W^2 + \tilde\pi_3(Y)\,W + \tilde\pi_4(Y).
\lab{PBWY_pol}
\ee

We can now state and prove the main result.

\begin{theorem}
\label{thm:elliptic}
Let $(X,Y)$ be a classical Leonard pair and let
\[
W = \tau_1 XY + \tau_2 Z + \tau_3 X + \tau_4 Y + \tau_0,
\qquad Z = \{X,Y\},
\]
be the associated classical Heun observable.  Then, on every energy surface
$W = \mathrm{const}$, the variables $X(t)$ and $Y(t)$ satisfy equations of
the form
\[
\dot X^2 = \mathcal{P}_4(X),
\qquad
\dot Y^2 = \widetilde{\mathcal{P}}_4(Y),
\]
where $\mathcal{P}_4$ and $\widetilde{\mathcal{P}}_4$ are polynomials of
degree at most four with time-independent coefficients.  In the generic
non-degenerate case, both $X(t)$ and $Y(t)$ are elliptic functions of
second order.
\end{theorem}

\begin{proof}
Fixing the energy $W = \mathrm{const}$ in~\eqref{PBWX_pol} expresses
$\dot X^2 = \{X,W\}^2$ as a polynomial $\mathcal{P}_4(X)$ of degree at most
four, with coefficients depending only on the parameters $\alpha_{ij}$,
$\tau_i$, and the fixed energy value.  Similarly,~\eqref{PBWY_pol} yields
$\dot Y^2 = \widetilde{\mathcal P}_4(Y)$.  For a generic quartic polynomial
\[
\mathcal P_4(x)
=
a\prod_{k=1}^{4}(x-e_k)
\]
with distinct roots, the solution is obtained by inversion of the
elliptic integral
\[
t-t_0=
\int_{x_0}^{x}
\frac{du}{\sqrt{\mathcal P_4(u)}}.
\]
The inverse of this integral is an elliptic function.
Since the quartic possesses four roots, the resulting solution
is an elliptic function of order two~\cite{Akhieser}.
\end{proof}

The non-generic cases in which the quartic degenerates
correspond to elementary or trigonometric limits of the
elliptic solution.

The explicit form of the solution is
\be
X(t) = \kappa\,\frac{\sigma(\mu(t-r_1))\,\sigma(\mu(t-r_2))}
                    {\sigma(\mu(t-r_3))\,\sigma(\mu(t-r_4))},
\lab{X_elliptic_1}
\ee
where $\sigma$ is the Weierstrass sigma function, $\kappa$, $\mu$, and the
poles $r_i$ are determined by the coefficients of $\mathcal{P}_4$ and the
initial condition $X(0)=X_0$, and satisfy the standard balance
condition~\cite{Akhieser}
\be
r_1 + r_2 = r_3 + r_4. \lab{balance}
\ee
The elliptic invariants $g_2, g_3$ (encoding the periods) are likewise
determined by the coefficients of $\mathcal{P}_4$.  The variables X and Y are parametrized by the same
elliptic curve determined by the fixed energy surface.

Theorem~\ref{thm:elliptic} may be interpreted as follows.  Choosing $X$ or
$Y$ alone as Hamiltonian (the Leonard case) produces second-degree polynomial
equations $\dot Y^2 = \mu_2(Y)$ or $\dot X^2 = \nu_2(X)$, whose solutions
are elementary.  Passing to the Heun deformation replaces the quadratic
right-hand side by a quartic, universally upgrading the dynamics from
elementary to elliptic.

\section{Applications}
\setcounter{equation}{0}
Theorem~\ref{thm:elliptic} is universal.
The examples below illustrate its realization in three
distinct algebraic settings:

\begin{itemize}
\item the classical Jacobi algebra;
\item the Lie--Poisson algebra $\mathfrak{su}(2)$;
\item the classical Askey--Wilson algebra.
\end{itemize}

In each case, the passage from a Leonard observable to its
Heun deformation transforms elementary dynamics into
elliptic dynamics.

\subsection{Extension of the P\"oschl--Teller Hamiltonian}

This first example is closest to Manning's original discussion~\cite{Manning}.
We show how ``elliptic'' potentials arise naturally as Heun deformations of
``elementary'' ones.

Recall the characterization of elementary potentials established
in~\cite{GLZ_Annals}.  Set $X = \varphi(q)$ for some function $\varphi$, and
let $Y = p^2 + u(q)$ be a one-particle Hamiltonian.  The canonical Poisson
bracket is
\be
\{q,p\} = 1. \lab{PB_qp}
\ee
The quadratic Jacobi algebra underlying elementary dynamics
(see~\cite{GLZ_Annals}) is equivalent to the condition $U_2(X) = 0$
in~\eqref{Phi_UV}, i.e.\
\be
Z^2 = \{X,Y\}^2 = \Phi(X,Y) = U_1(X)\,Y + U_0(X), \lab{Z2_Jacobi}
\ee
which amounts to the two conditions
\be
4\,\varphi'(q)^2 = U_1(\varphi(q)) \lab{1_con_J}
\ee
and
\be
u(q) = -\frac{U_0(\varphi(q))}{U_1(\varphi(q))}. \lab{2_con_J}
\ee
When $U_1$ is a quadratic polynomial with distinct roots,
equation~\eqref{1_con_J} determines $\varphi$ as a trigonometric or
hyperbolic function.  The canonical choice is
\be
\varphi(q) = \sinh^2 q. \lab{phi_sin}
\ee
Equation~\eqref{2_con_J} then yields the generic P\"oschl--Teller
potential~\cite{Manning, GLZ_Annals}
\be
u(q) = \beta_1 \sinh^{-2} q + \beta_2 \cosh^{-2} q + \beta_0,
\lab{u_Jac}
\ee
and the variable $X(t) = \sinh^2 q(t)$ evolves as an elementary function of
time.

We now form the Heun pencil~\eqref{cl_W} with $\tau_4 = 1$ and $\tau_1 = 0$,
leaving three free parameters $\tau_0, \tau_2, \tau_3$.  After the canonical
shift $p \mapsto p - \tau_2 \varphi'(q)$, which eliminates the cross-term
$\tau_2 Z$ at the cost of modifying the potential, the Hamiltonian takes the
form
\be
W = p^2 + \beta_1 \sinh^{-2} q + \beta_2 \cosh^{-2} q
    + \beta_3 \sinh^2 q + \beta_4 \sinh^2 q \cosh^2 q + \beta_0,
\lab{W_PTE}
\ee
where $\beta_0 = \tau_0$, $\beta_3 = -\tau_3$, and $\beta_4$ is proportional
to $\tau_2^2$, while $\beta_1, \beta_2$ are the original P\"oschl--Teller
parameters.  This is a five-parameter extension of the P\"oschl--Teller
potential. By
Theorem~\ref{thm:elliptic}, the variable $X(t) = \sinh^2 q(t)$ evolves as an
elliptic function of second order for any values of the parameters
$\beta_i$.  This provides an algebraic mechanism underlying Manning's
observation: a Heun deformation of a system governed by
the classical Jacobi algebra naturally produces elliptic
classical dynamics.

\begin{remark}
The class of potentials for which the Hamilton--Jacobi equation admits an
``elliptic'' form is strictly larger than the class for which the
\emph{time} dynamics is elliptic.  The Heun pencil approach captures precisely
the latter.
\end{remark}

\subsection{The Zhukovsky--Volterra gyrostat}

The second example concerns the classical Leonard pair on the Lie--Poisson
algebra $\mathfrak{su}(2)$, defined by the standard brackets
\be
\{s_i, s_k\} = \varepsilon_{ikl}\,s_l, \quad i,k,l = 1,2,3, \lab{PB_su(2)}
\ee
where $\varepsilon_{ikl}$ is the Levi-Civita symbol.  The Casimir element
\be
S^2 = s_1^2 + s_2^2 + s_3^2 \lab{S_su(2)}
\ee
satisfies $\{S^2, s_i\} = 0$ and is fixed as a positive real constant $S > 0$.

We choose the classical Leonard pair
\be
X = s_1 + \beta s_2, \quad Y = s_1 - \beta s_2, \lab{XY_su(2)}
\ee
with real parameter $\beta$.  A direct computation gives
\be
Z = \{X,Y\} = -2\beta s_3 \lab{Z_su(2)}
\ee
and
\be
Z^2 = 4\beta^2 s_3^2
    = \Phi(X,Y)
    = 4S^2\beta^2 - (\beta^2+1)(X^2+Y^2) + 2(1-\beta^2)XY,
\lab{Z2_su2}
\ee
so $\Phi$ has total degree two and $(X,Y)$ is indeed a CLP.

Setting $\tau_0 = 0$, the Heun pencil~\eqref{cl_W} reads
\be
W = \tau_1(s_1^2 - \beta^2 s_2^2) - 2\beta\tau_2 s_3
    + (\tau_3+\tau_4) s_1 + \beta(\tau_3-\tau_4) s_2. \lab{W_su2}
\ee
The leading term $s_1^2 - \beta^2 s_2^2$ is the Euler-top Hamiltonian; the
remaining terms constitute a linear perturbation.  The Hamiltonian~\eqref{W_su2}
is precisely the \emph{Zhukovsky--Volterra gyrostat}~\cite{Basak}, a
completely integrable rigid-body model.

By Theorem~\ref{thm:elliptic}, both $X(t)$ and $Y(t)$ --- and hence the
individual spin components --- are elliptic functions of second order.  The
classical Heun pencil thus provides a transparent algebraic derivation of the
integrability of the Zhukovsky--Volterra gyrostat and identifies, in a
systematic way, which linear combinations of $s_1, s_2, s_3$ evolve
elliptically.  This identification is considerably less transparent in direct
approaches; see the analysis in~\cite{Basak}.

The quantum analog of this construction, leading to a quantum gyrostat, was
studied in~\cite{Crampe}.

\subsection{Extension of the relativistic $A_1$ model}

The third example provides a \textit{dequantization} of the Askey--Wilson
polynomials and is related to the classical $A_1$ Ruijsenaars model~\cite{Ruijs}.

The Hamiltonian is taken to be
\be
Y = u(q)\cosh p, \lab{Y_A1}
\ee
where $\cosh p$ is the relativistic kinetic energy and $u(q)$ a relativistic
potential.  Setting $X = \varphi(q)$ as before, we seek conditions under
which $(X,Y)$ forms a classical Leonard pair in the sense of the Askey--Wilson
algebra (cf.~\cite{ZK}).

The fundamental identity~\eqref{Z^2_Phi} requires
\be
Z^2 = \{X,Y\}^2 = U_2(X)\,Y^2 + U_0(X), \lab{rel_NC}
\ee
with arbitrary second-degree polynomials $U_2(X)$ and $U_0(X)$.  (The term
$U_1(X)\,Y$ vanishes by the specific structure of~$Y$.)  Since
\be
Z = u(q)\,\varphi'(q)\sinh p, \lab{Z_A1}
\ee
condition~\eqref{rel_NC} reduces to
\be
\varphi'(q)^2 = U_2(\varphi(q)) \lab{phi_eq_r}
\ee
and
\be
u^2(q) = -\frac{U_0(\varphi(q))}{U_2(\varphi(q))}. \lab{u_rel}
\ee
Equation~\eqref{phi_eq_r} is of the same form as~\eqref{1_con_J}, so the
non-degenerate solution is again $\varphi(q) = \sinh^2 q$.  Equation~\eqref{u_rel}
then gives the relativistic P\"oschl--Teller potential
\be
u^2(q) = \beta_1 \sinh^{-2} q + \beta_2 \cosh^{-2} q + \beta_0,
\lab{u2_rel}
\ee
which for $\beta_2 = 0$ specializes to the Ruijsenaars $A_1$
potential~\cite{Ruijs}.  With these choices, $X = \sinh^2 q$ evolves as an
elementary function of time --- a fact that now has a purely algebraic
explanation through the CLP structure.

The Heun pencil associated with this model is
\be
W = (\tau_1 \sinh^2 q + \tau_4)\,u(q)\cosh p
    + \tau_2\,u(q)\,\varphi'(q)\sinh p
    + \tau_3 \sinh^2 q. \lab{W_rel}
\ee
The term involving $\sinh p$ can be removed by the canonical
transformation $p \mapsto p + \chi(q)$ for an appropriate $\chi(q)$,
reducing $W$ to the simpler form
\be
W = \Phi_1(q)\cosh p + \Phi_0(q), \lab{W_Phi}
\ee
where
\be
\Phi_0(q) = \tau_3 \sinh^2 q, \lab{Phi0}
\ee
\be
\Phi_1(q) = u(q)\sqrt{(\tau_1 \sinh^2 q + \tau_4)^2
            - \tau_2^2 \sinh^2 q\,}. \lab{Phi_12}
\ee
By Theorem~\ref{thm:elliptic}, the variable $X(t) = \sinh^2 q(t)$ is an
elliptic function of second order for the Hamiltonian~\eqref{W_rel} with
arbitrary parameters $\tau_i$.

\subsection{Semiclassical quantization of space}

Bianchi and Haggard proposed a semi-classical approach to quantum gravity. One of the most important object in this approach is classical Hamiltonian chosen as the volume of tetrahedron constructed from three classical vectors $\bf A, \bf B, \bf C$ subjected to classical rotation algebra $so(3)$ relations:
\be
\{\bf A, \bf B\} = \bf C
\lab{PB_AB} \ee
(and two other relations by cyclic permutation of vectors $\bf A, \bf B, \bf C$).

Take the following quantities 
\be
K_1 = (\bf A + \bf B)^2,  \; K_2 = (\bf B + \bf C)^2
\lab{gen_R} \ee
as basic generators of (classical) Racah algebra. 

The quantities $\bf A^2, \bf B^2, \bf C^2, (\bf A +\bf B + \bf C)^2$ have zero Poisson brackets with $K_1, K_2$ and thus can be chosen as constants.
The classical Racah algebra behind $K_, K_2$ has the form
\ba
&&\{ K_1, K_2\} = K_3,  \nonumber \\
&&\{ K_3, K_1\} = \alpha_1 K_1^2 + 2 \alpha_2 K_1K_2 + \alpha_3 K_1 + \alpha_4 K_2 + \alpha_5,  \lab{Racah}  \\
&& \{ K_2, K_3\} = 2 \alpha_1 K_1 K_2 +  \alpha_2 K_2^2 + \alpha_3 K_2 + \alpha_6 K_1 + \alpha_7, \nonumber \ea
where
\be
K_3 = 4 V 
\lab{K_2_V} \ee
and where 
\be
V=\left(\bf A, \left[\bf B , \bf C\right] \right)
\lab{V} \ee
is (directed) volume of the parallelepiped constructed out of vectors $\bf A, \bf B, \bf C$. The structure parameters $\alpha_i, i=1,\dots, 7$ depend on the four constants $\bf A^2, \bf B^2, \bf C^2, (\bf A +\bf B + \bf C)^2$ (we need not concrete expressions of these parameters).

By previous considerations, we can construct Heun Hamiltonian
\be
H =  \tau_1 K_3 + \tau_2 K_1 K_2 + \tau_3 K_1 + \tau_4 K_2
\lab{Heun_R} \ee

\section{Conclusions}
\setcounter{equation}{0}

We have identified a universal algebraic mechanism that produces elliptic
dynamics from elementary dynamics.  Starting from a classical Leonard pair
$(X,Y)$ --- whose mutual integrability is encoded in second-degree polynomial
equations for $\dot X^2$ and $\dot Y^2$ --- we form the associated classical
Heun observable and show that it generates Hamiltonian flows governed by
\emph{quartic} polynomial equations.  The distinguished observables of the
Leonard pair consequently evolve as elliptic functions of second order. 

In this sense, the classical Heun observable plays for
elliptic solvability the same role that the classical
Leonard pair plays for elementary solvability.

From the bispectral perspective, the passage from a Leonard observable to its
Heun deformation increases the complexity of the underlying dynamics from
elementary to elliptic.  This phenomenon is independent of the specific
realization of the classical Leonard pair --- whether on a cotangent bundle, a
Lie--Poisson algebra, or a relativistic phase space --- and therefore
constitutes a universal feature of classical Heun observables.

Our results provide an algebraic explanation of the second
part of Manning's observation. They also reveal a reverse
mechanism within the present algebraic framework:
the Heun deformation of a classically solvable system
with elementary dynamics generates elliptic dynamics.
The three worked examples ---
the extended P\"oschl--Teller system, the Zhukovsky--Volterra gyrostat, and
the relativistic $A_1$ model --- illustrate the breadth of the framework.

Several natural directions remain open.  It would be interesting to classify
which degenerate cases of the quartic $\mathcal{P}_4$ arise from specific
families of classical Leonard pairs, to investigate higher-rank
generalizations, and to explore the quantization of the classical Heun
observables constructed here.

\bigskip\bigskip
{\bf \Large Acknowledgments}

\noindent
The work of L.V.\ is supported in part by the Natural Sciences and
Engineering Research Council (NSERC) of Canada.  A.Z.\ is supported by the
Ministry of Science and Higher Education of the Russian Federation (agreement
no.\ 075--15--2025--343).

\subsection*{Conflict of interest}
On behalf of all authors, the corresponding author states that there is no
conflict of interest.

\subsection*{Data availability}
The authors declare that all data supporting the findings of this study are
contained within the paper.

\newpage

\bb{99}

\bi{Akhieser} Akhiezer, N.I., \textit{Elements of the Theory of Elliptic
Functions}, 2nd edn.\ Nauka, Moscow (1970); English transl., Translations in
Mathematics Monographs, vol.\ \textbf{79}, Amer.\ Math.\ Soc., Providence
(1990).

\bi{Basak} Basak, I., Explicit solution of the Zhukovski--Volterra gyrostat.
\textit{Regular and Chaotic Dynamics} \textbf{14} (2009), 223--236.

\bi{BH} E. Bianchi, H.M. Haggard, {\it Bohr-Sommerfeld quantization of space}, Phys. Rev. {\bf D 86}, 124010 (2012).

\bi{Crampe} Cramp\'e, N., Vinet, L., Zhedanov, A., Heun algebras of Lie type.
\textit{Proc.\ Amer.\ Math.\ Soc.} \textbf{148} (2020), 1079--1094.
ArXiv:1904.10643.

\bi{GLZ_Annals} Granovskii, Ya.\,A., Lutzenko, I.M., Zhedanov, A., Mutual
integrability, quadratic algebras, and dynamical symmetry.
\textit{Ann.\ Phys.} \textbf{217} (1992), 1--20.

\bi{GVZ_Heun} Gr\"unbaum, F.A., Vinet, L., Zhedanov, A., Tridiagonalization
and the Heun equation.
\textit{J.\ Math.\ Phys.} \textbf{58} (2017), 031703. ArXiv:1602.04840.

\bi{GVZ_band} Gr\"unbaum, F.A., Vinet, L., Zhedanov, A., Algebraic Heun
operator and band--time limiting.
\textit{Comm.\ Math.\ Phys.} \textbf{364} (2018), 1041--1068.
ArXiv:1711.07862.

\bi{Manning} Manning, M., Exact solutions of the Schr\"odinger equation.
\textit{Phys.\ Rev.} \textbf{48} (1935), 161--164.

\bi{Ruijs} Ruijsenaars, S., The classical hyperbolic Askey--Wilson dynamics
without bound states.
\textit{Theor.\ Math.\ Phys.} \textbf{154}(3) (2008), 418--432.

\bi{Tur_Heun} Turbiner, A., The Heun operator as a Hamiltonian.
\textit{J.\ Phys.\ A} \textbf{49} (2016), 26LT01. ArXiv:1603.02053.

\bi{Tur_quasi} Turbiner, A., One-dimensional quasi-exactly solvable
Schr\"odinger equations.
\textit{Phys.\ Rep.} \textbf{642} (2016), 1--71. ArXiv:1603.02992.

\bi{ZK} Zhedanov, A., Korovnichenko, A., Leonard pairs in classical mechanics.
\textit{J.\ Phys.\ A: Math.\ Gen.} \textbf{35} (2002), 5767--5780.

\eb

\end{document}